\documentclass[aps,prl,twocolumn,superscriptaddress]{revtex4-2}
\usepackage[T1]{fontenc}
\usepackage[utf8]{inputenc}
\usepackage{graphicx}
\usepackage{upgreek}
\usepackage{color}
\usepackage[a4paper,
colorlinks=true,
linkcolor=blue,
citecolor=blue,
urlcolor = blue]{hyperref}

\begin{document}


\title{Three-dimensional structure and formation mechanism of biskyrmions in uniaxial ferromagnets}



\author{Cheng-Jie Wang}
\affiliation{CAS Key Laboratory of Microscale Magnetic Resonance and School of Physical Sciences, University of Science and Technology of China, Hefei~230026, China}
\affiliation{CAS Center for Excellence in Quantum Information and Quantum Physics, University of Science and Technology of China, Hefei~230026, China}

\author{Pengfei Wang}
\email{wpf@ustc.edu.cn}
\affiliation{CAS Key Laboratory of Microscale Magnetic Resonance and School of Physical Sciences, University of Science and Technology of China, Hefei~230026, China}
\affiliation{CAS Center for Excellence in Quantum Information and Quantum Physics, University of Science and Technology of China, Hefei~230026, China}

\author{Yan Zhou}
\email{zhouyan@cuhk.edu.cn}
\affiliation{School of Science and Engineering, The Chinese University of Hong Kong, Shenzhen, Guangdong 518172, China}

\author{Wenhong Wang}
\affiliation{Beijing National Laboratory for Condensed Matter Physics and Institute of Physics, Chinese Academy of Sciences, Beijing 100190, China}

\author{Fazhan Shi}
\affiliation{CAS Key Laboratory of Microscale Magnetic Resonance and School of Physical Sciences, University of Science and Technology of China, Hefei~230026, China}
\affiliation{CAS Center for Excellence in Quantum Information and Quantum Physics, University of Science and Technology of China, Hefei~230026, China}

\author{Jiangfeng Du}
\email{djf@ustc.edu.cn}
\affiliation{CAS Key Laboratory of Microscale Magnetic Resonance and School of Physical Sciences, University of Science and Technology of China, Hefei~230026, China}
\affiliation{CAS Center for Excellence in Quantum Information and Quantum Physics, University of Science and Technology of China, Hefei~230026, China}	


\date{\today}

\begin{abstract}
Magnetic biskyrmions are observed in experiments but their existences are still under debate. In this work, we present the existence of biskyrmions in a magnetic film with tilted uniaxial anisotropy via micromagnetic simulations. We find biskyrmions and bubbles share a unified three-dimensional structure, in which the relative position of two intrinsic Bloch points dominates the two-dimensional topological property in the film middle. The film edge can drive Bloch points and transform bubbles into biskyrmions via the demagnetizing field. This mechanism is found in the formation process of biskyrmions in confined geometry under zero field. Our work clarifies the structure and formation mechanism of biskyrmions, emphasizing the three-dimensional aspect of skyrmion-related nanostructures.
\end{abstract}

\maketitle 


Magnetic skyrmions are particle-like spin textures of nontrivial topology found in many materials \cite{nagaosa2013topological,tokura_magnetic_2021}. It has received much attention due to both fundamental interests in topological matters and potential applications in spintronics \cite{iwasaki_current-induced_2013,xing_skyrmion_2016,pinna_skyrmion_2018,song_skyrmion-based_2020}. While intensive efforts have been made to improve the applicability of conventional skyrmions, alternative topological objects have been studied recently \cite{nayak_magnetic_2017,zhang_real-space_2018,gao_creation_2019,gobel_beyond_2021}. These spin textures can exhibit different properties and bring new possibilities for innovative device applications \cite{zheng_experimental_2018,gobel_electrical_2019,jena_evolution_2020,zarzuela_stability_2020}. Among them, biskyrmions observed in centrosymmetric magnets \cite{yu_biskyrmion_2014,wang_centrosymmetric_2016} attract lots of attention due to its rare topological charge of two even in the absence of Dzyaloshinskii-Moriya interaction (DMI). However, it is later proposed that biskyrmions could be misleading images of topologically trivial bubbles observed by Lorentz transmission electron microscopy (TEM) \cite{loudon_images_2019,yao_magnetic_2019,chen_effects_2021} since biskyrmions have not been considered under realistic conditions as in experiments.

Nonetheless, magnetic bubbles cannot explain the observation of the large topological Hall effect (THE) \cite{wang_centrosymmetric_2016}, although THE is not sufficient evidence for the existence of skyrmions. In addition, the stability of biskyrmions in films of finite thickness has been confirmed theoretically \cite{capic_stabilty_2019}. Therefore, the existence of biskyrmions still remains an open problem to be solved. Alternative techniques have to be utilized to provide more experimental evidence. On the other hand, it is imperative to understand the formation mechanism of biskyrmions theoretically which can serve as guides of realizing and utilizing biskyrmions in the experiments. The three-dimensional (3D) structure of biskyrmions could provide the key to addressing this issue since both the bubble explanation and the stability analysis are 3D in nature. In fact, the 3D extensions of skyrmions have become an emerging area to explore \cite{milde_unwinding_2013,birch_real-space_2020,kanazawa_critical_2016,kanazawa_direct_2020}, partially because of the finite-thickness nature of skyrmions. Especially, a new type of 3D topological spin texture called hopfions has been proposed and studied recently \cite{ackerman_static_2017,tai_topological_2018,kent_creation_2021}.

In this work, we present the three-dimensional structure of magnetic biskyrmions in uniaxial ferromagnets and analyze its formation mechanism based on micromagnetic simulation. We simulate a thin film with the easy axis at a tilted angle to the surface normal and realize biskyrmions in the film middle. The 3D structure of biskyrmions is similar to type-II bubbles \cite{rothman_observation_2001,phatak_nanoscale_2016}, in which the magnetization points radially near the surface \cite{loudon_images_2019}. In the film middle, the difference between topologically trivial bubbles and biskyrmions is dominated by the relative position of two intrinsic Bloch points (BPs) \cite{im_dynamics_2019} in the thickness direction. BPs can be driven across each other in the thickness direction by film edges, inducing the formation of biskyrmions. This mechanism can be utilized to generate biskyrmions in a stripe-shape magnetic film.

\begin{figure*}
	\includegraphics[scale=0.192]{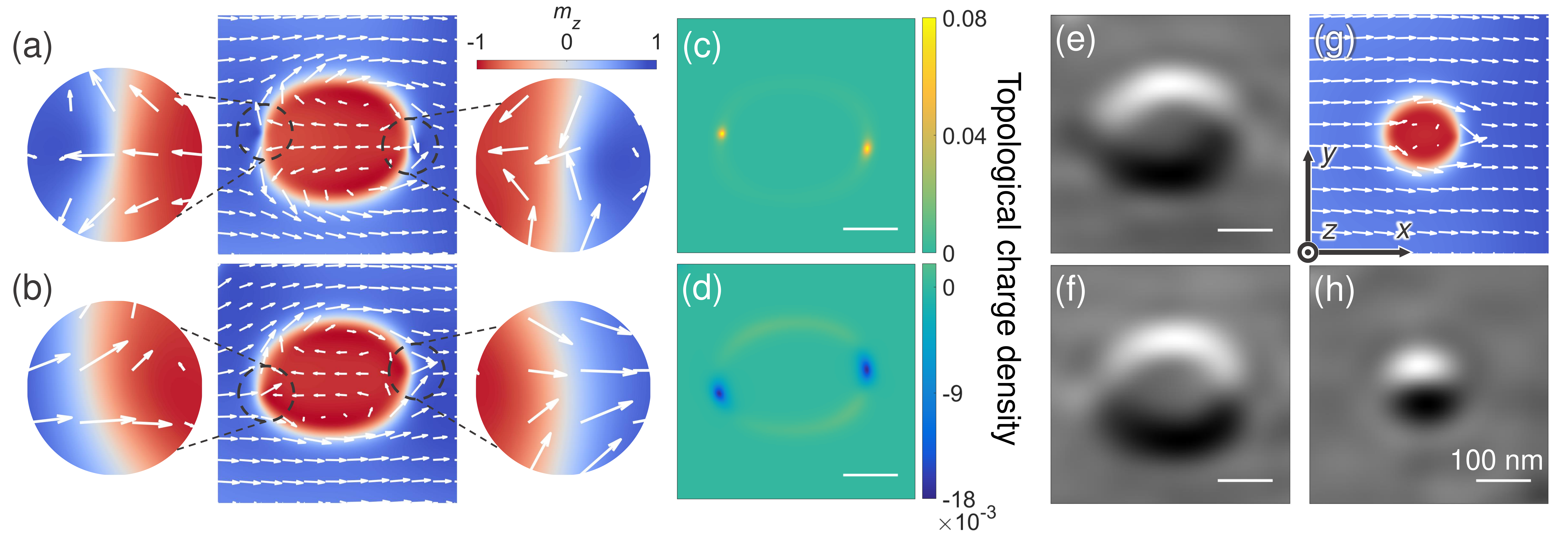}
	\caption{(a)(b) The spin texture of the biskyrmion slightly below the film middle and the bubble in the film middle, respectively.  The arrow represents the in-plane magnetization $m_{xy}$. (c)(d) The distribution of topological charge density, i.e., the integrand in Eq.\ref{eq:Q}, of spin textures in (a) and (b), respectively. (g) The bubble transformed from (a) at 0.2 $\mathrm{T}$. The external field is increased with steps of 0.02 $\mathrm{T}$, and the state is relaxed in every step. The magnetic easy axis lies in xz-plane. (e)(f)(h) Corresponding Lorentz TEM images of (a)(b)(g), respectively. Calculation is performed with magnetization averaged through the thickness at defocus of 1.6 mm \cite{walton_malts_2013}.}\label{fig:sim}
\end{figure*}

Micromagnetic simulations in this work are performed using Mumax3 \cite{vansteenkiste_design_2014}, considering the exchange, dipolar, uniaxial anisotropy and Zeeman interactions. The material parameters of $ \mathrm{MnNiGa} $ alloy are listed as follows \cite{loudon_images_2019}: exchange stiffness $ A=2\times10^{-11}~\mathrm{J}\cdot\mathrm{m}^{-1}$, saturation magnetization $  M_s=5.16\times10^5~\mathrm{A}\cdot\mathrm{m}^{-1}$, and uniaxial anisotropy $ K_u=8.7\times10^4~\mathrm{J}\cdot\mathrm{m}^{-3}$. Previous studies report that the crystal orientation, thus the magnetic easy axis direction, is a key ingredient to form biskyrmions \cite{ding_crystal-orientation_2018}. Hence, we perform simulations with the easy axis at a tilted angle of $ 31^\circ $ as in the $ [12\bar{2}] $ crystal \cite{ding_crystal-orientation_2018}. The thickness of films in simulations is 160 $\mathrm{nm}$ to constrain the formation of type-I bubbles \cite{loudon_images_2019}, corresponding to the typical film thickness in experiments. The values of geometry, cell size, and external field can be varied and are always specified in figure captions or the main text.

The simulation is initialized by relaxing a state with random magnetization orientation at an external magnetic field normal to the sample. We start with a sample of $1.2~\upmu\mathrm{m}\times 1.2~\upmu\mathrm{m} \times160~\mathrm{nm}$  size with cells of $2~\mathrm{nm}\times 2~\mathrm{nm} \times4~\mathrm{nm}$ at an external field of 0.12 $\mathrm{T}$. It is known that two types of magnetic bubbles can emerge in the uniaxial magnetic film. Furthermore, the presence of tilted easy axis can introduce in-plane anisotropy so that type-II bubbles become energetically favorable. As a result, type-II bubbles are most commonly observed in our simulations \cite{SM}. In contrast with the spin texture of a typical bubble shown in Fig.\ref{fig:sim}(b), a biskyrmion can be found in Fig.\ref{fig:sim}(a). Although the in-plane anisotropy causes in-plane magnetization inside and outside the domain wall in both bubbles and biskyrmions, it has no influence on their topological nature. The topological charge of these spin textures can be calculated as:
\begin{equation}
	Q=1/4\uppi\int \mathrm{d}x\mathrm{d}y~{\mathbf{m}}\cdot \partial_x {\mathbf{m}}\times \partial_y{\mathbf{m}}
	\label{eq:Q}
\end{equation}
where ${\mathbf{m}}$ is the unit vector of magnetization. The topological charge $ Q $ of the spin texture is 2 in Fig.\ref{fig:sim}(a) against 0 in Fig.\ref{fig:sim}(b) via numerical calculation. Nevertheless, the spin texture in Fig.\ref{fig:sim}(a) is not the very model assumed with Lorentz TEM images, in which a biskyrmion is composed of two Bloch-type skyrmions with opposite helicities. By contrast, it is a thin-wall biskyrmion bubble, whereas it would still be called a biskyrmion to be distinguished from topologically trivial bubbles in the following.

We note that the structures in Fig.\ref{fig:sim}(a) and Fig.\ref{fig:sim}(b) are quite similar. The main difference is the direction of Bloch lines in the domain wall indicated by the dashed circles, which turns out to be the key factor for the topological charge as indicated in Fig.\ref{fig:sim}(c) and Fig.\ref{fig:sim}(d). The reason can be better understood by the winding number associated with domain walls, which describes a $ 2\uppi $ twist along the domain wall \cite{cheng_magnetic_2019,kuchkin_magnetic_2020}. Although this interpretation is equivalent to the topological charge in Eq. \ref{eq:Q}, it emphasizes the structure of domain walls. In detail, the Bloch lines reverse the magnetization twice and thus make a twist of $ 4\uppi $ along the domain wall in Fig.\ref{fig:sim}(a), whereas the opposite Bloch lines cancel the twist along the wall in Fig.\ref{fig:sim}(b). When the external field is up to 0.2 $\mathrm{T}$, the biskyrmion in Fig.\ref{fig:sim}(a) becomes a standard type-II bubble, as shown in Fig.\ref{fig:sim}(g). It should be noted that bubbles in Fig.\ref{fig:sim}(b) can also become more standard as in Fig.\ref{fig:sim}(g) under a high field.

To compare with the previous experimental results, we calculate the Lorentz TEM images of these spin textures displayed in Fig.\ref{fig:sim}. As can be seen, images in Fig.\ref{fig:sim}(e) and Fig.\ref{fig:sim}(f) are similar to each other and consistent with previous experimental results at low external field after field cooling \cite{peng_real-space_2017}. Moreover, the image shown in Fig.\ref{fig:sim}(h) consisting of black and white semicircles is in good agreement with experimental results observed at high field \cite{yu_biskyrmion_2014,wang_centrosymmetric_2016}. These results confirm that tilted bubbles can produce the Lorentz TEM images like biskyrmions, as reported previously \cite{loudon_images_2019,chen_effects_2021}, but the spin texture corresponding to the image could still be a biskyrmion.

\begin{figure}
	\includegraphics[scale=0.45]{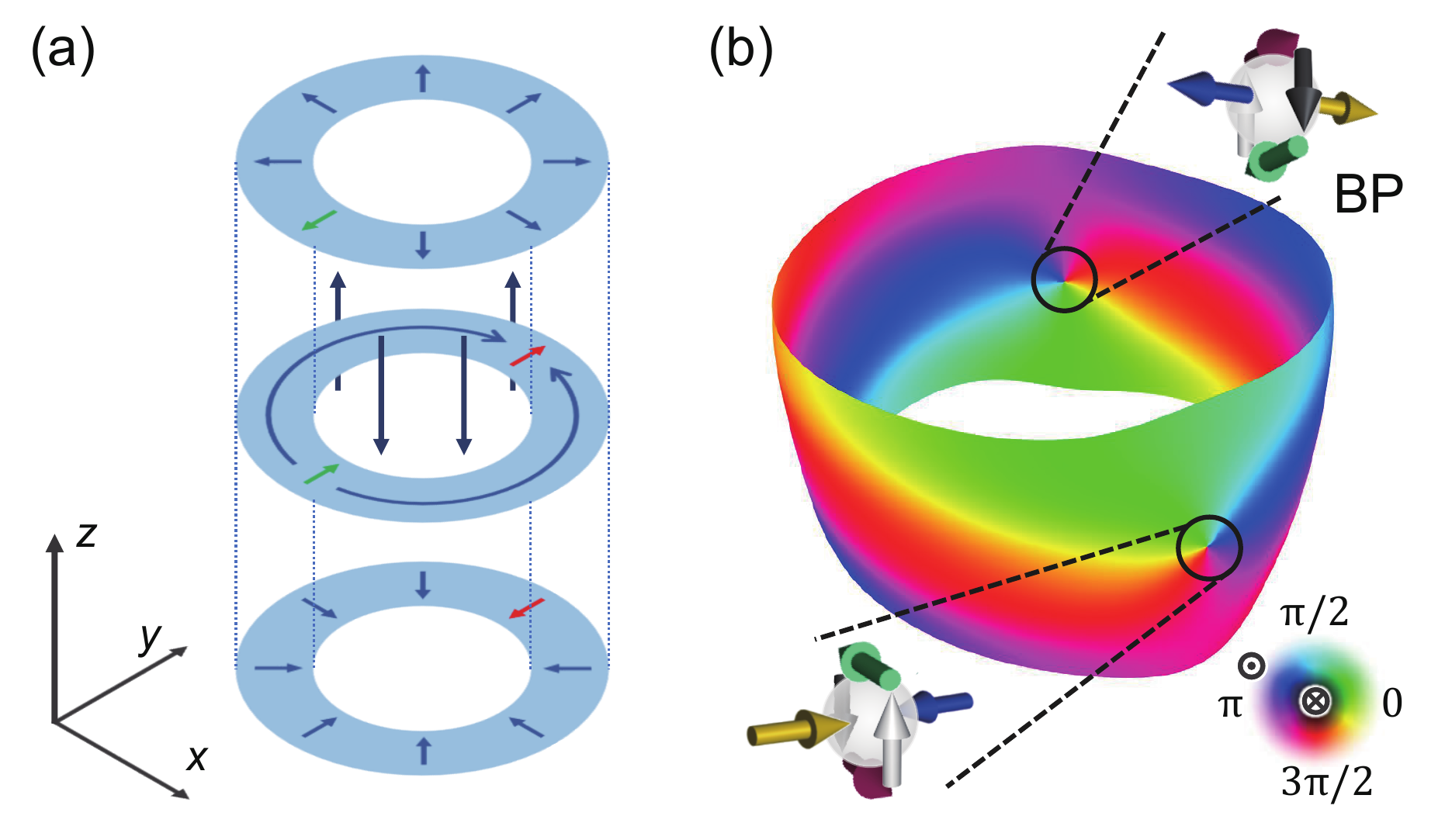}
	\caption{(a) Illustration of the 3D structure of a type-II bubble. (b) The contour surface of $ {\mathbf{m}}_z=0 $  of a simulated bubble and diagrams of two BPs on it. The sample is defined as  $2~\upmu\mathrm{m}\times 2~\upmu\mathrm{m} \times200~\mathrm{nm}$  size with cells of $4~\mathrm{nm}\times 4~\mathrm{nm} \times4~\mathrm{nm}$. The simulation is performed with perpendicular easy axis at a perpendicular external field of 0.06 T. Other parameters are the same as in the main text. The color represents the direction of magnetizations as conventional color wheel at the right bottom corner. The contour surfaces in this work are drawn with the software Spirit \cite{muller_spirit_2019}.}\label{fig:bubble}
\end{figure}

Next, we inspect the 3D spin structures to explore the formation mechanism of biskyrmions. Although skyrmions are usually assumed to be homogeneous through the film thickness, the surface magnetization of a Bloch-type skyrmion can form N$\mathrm{\acute{e}}$el-caps \cite{montoya2017tailoring,dovzhenko_magnetostatic_2018,legrand_hybrid_2018}. Such configurations also exist in the 3D structure of bubbles \cite{loudon_images_2019}. Therefore, we revisit the 3D structure of the type-II bubble as shown in Fig.\ref{fig:bubble}(a). As can be seen, the magnetization forms a type-II bubble in the middle while it points out (in) radially at the top (bottom). The magnetization indicated by green (red) arrows is opposite between the middle and the top (bottom). An intrinsic 3D BP resides between them, considering the magnetization is also opposite inside and outside the domain wall. The contour surface of $ {\mathbf{m}}_z=0 $ of a standard type-II bubble, i.e., the two-dimensional domain wall, is depicted in Fig.\ref{fig:bubble}(b). Two BPs exist on this surface, as indicated by the circles in Fig.\ref{fig:bubble}(b).

A BP can be characterized by a 3D skyrmion charge \cite{im_dynamics_2019}:
\begin{equation}
q=1/8\uppi\int \mathrm{d}A_i\epsilon_{ijk}~{\mathbf{m}}\cdot \partial_j {\mathbf{m}}\times \partial_k{\mathbf{m}}
\end{equation}
where the integration is taken over a closed surface $ A $ surrounding the BP. According to this, the two BPs in Fig.\ref{fig:bubble}(b) have opposite charges.

\begin{figure}
	\includegraphics[scale=0.38]{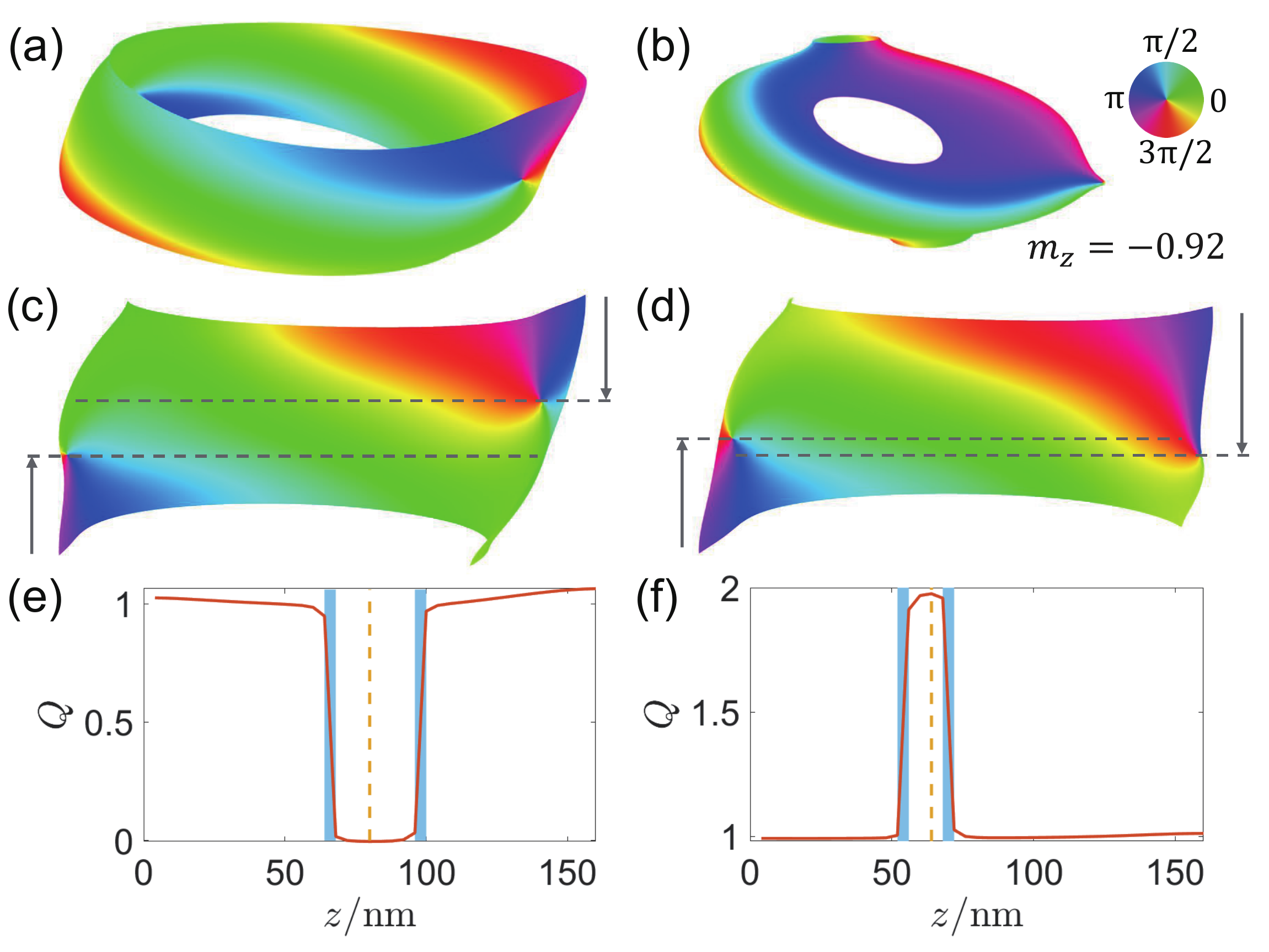}
	\caption{(a)(b) Contour surfaces of $ \mathbf{m}_z=0$ and $\mathbf{m}_z=-0.92$ of the bubble in Fig.\ref{fig:sim}(b). The color represents the direction of magnetization without z-component as the color wheel at the right top corner. (c)(d) The half of the surface $\mathbf{m}_z=0 $ of the bubble in Fig.\ref{fig:sim}(b) and the biskyrmion in Fig.\ref{fig:sim}(a), respectively. (e)(f) The dependence of the topological charge $ Q  $ with perpendicular position $ z $ of the bubble in Fig.\ref{fig:sim}(b) and the biskyrmion in Fig.\ref{fig:sim}(a), respectively. The perpendicular positions of slices in Fig.\ref{fig:sim}(b) and Fig.\ref{fig:sim}(a) is indicated by orange dashed lines.}\label{fig:BP}
\end{figure}

BPs appear to be roughly aligned along the easy direction in the presence of the tilted anisotropy and the region magnetized along the easy axis (green color in Fig.\ref{fig:BP}) is enlarged, as shown in Fig.\ref{fig:BP}(a). Moreover, it is interesting to note that the 3D structure of the core of bubbles or biskyrmions is topologically different from a tube, as demonstrated in Fig.\ref{fig:BP}(b). 

 The 3D structure of biskyrmions is quite similar with bubbles. Fig.\ref{fig:BP}(c) and Fig.\ref{fig:BP}(d) demonstrate the half of the surface $ {\mathbf{m}}_z=0 $ of the bubble in Fig.\ref{fig:sim}(b) and the biskyrmion in Fig.\ref{fig:sim}(a), respectively. Comparing Fig.\ref{fig:BP}(c) with Fig.\ref{fig:BP}(d) shows that the major difference between a bubble and a biskyrmion is the relative position of BPs. In other words, the BPs in biskyrmions are across each other in the thickness direction. Specifically, the formation of the biskyrmion in Fig.\ref{fig:sim}(a) is induced by the motion of the upper BP, which is clearly illustrated by Fig.\ref{fig:BP}(d). 

These findings can also be confirmed by the dependence of the topological charge $ Q $ with perpendicular position $ z $, as shown in Fig.\ref{fig:BP}(e) and Fig.\ref{fig:BP}(f). Because of N$\mathrm{\acute{e}}$el-caps on the surfaces, topological charges of both bubbles and biskyrmions are equal to 1 near the surfaces. Such structures along the thickness thus can dominate the measurement of THE. As a result, even topologically trivial bubbles can exhibit THE. Meanwhile, the difference in topology between bubbles and biskyrmions only exists in the middle. Therefore, the topological charge changes in two transition positions through the thickness. A change of topological charge through the thickness has to be attributed to a BP. As can be seen, the transition positions of topological charges agree with the locations of BPs depicted by blue shades, which provides compelling evidence for the BPs' role in the formation of biskyrmions.

In addition, it can be inferred that the motion of BPs accounts for the transition from biskyrmions to bubbles derived from the increase of the external field. Increasing the field drives BPs towards the surface and expands the zero-$ Q $ thickness. In summary, biskyrmions and bubbles have different topologies as a two-dimensional structure, while they share a unified 3D model. This result emphasizes the significance of the 3D aspect of topological objects in magnetism.

\begin{figure}
	\includegraphics[scale=0.5]{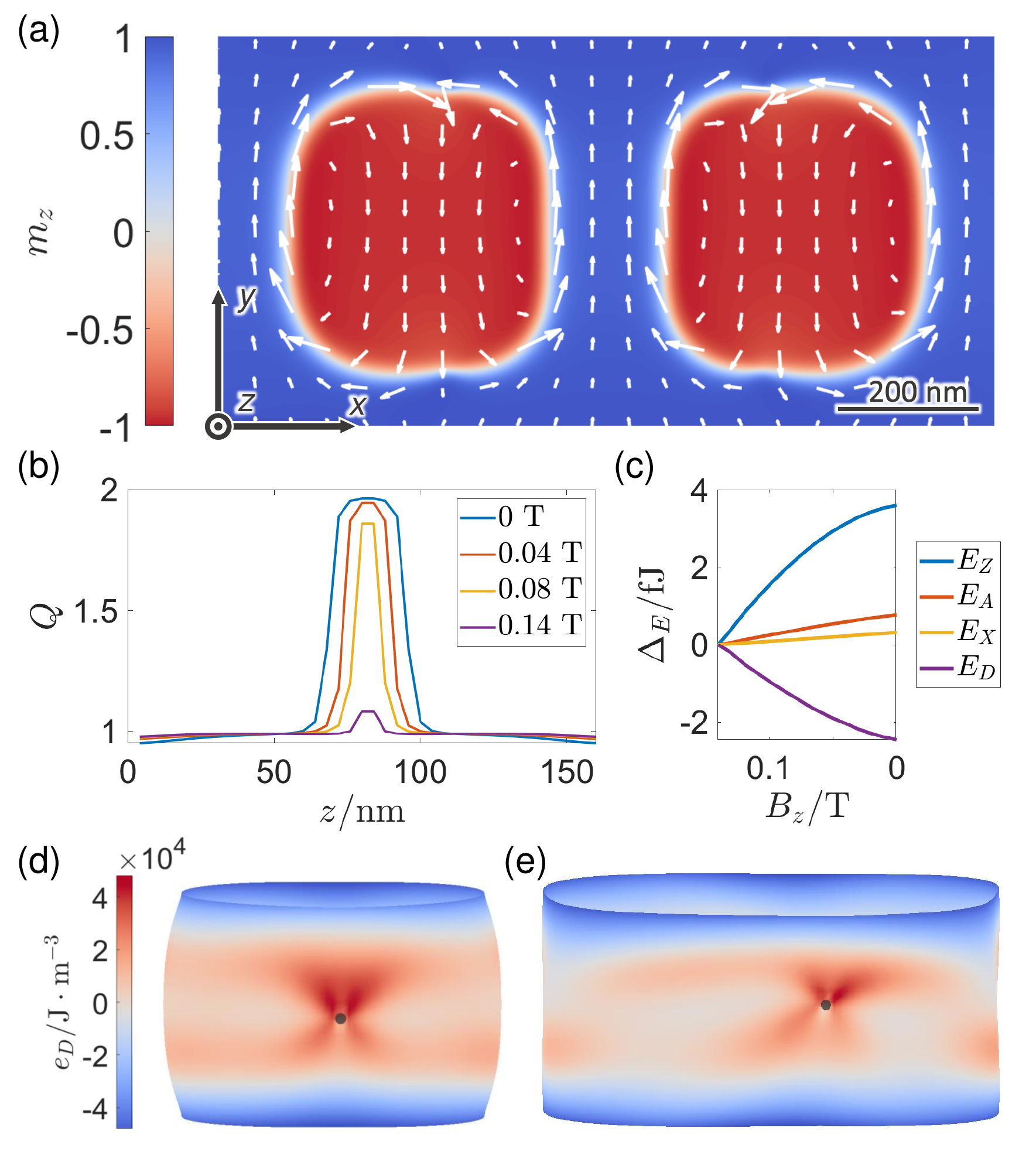}
	\caption{(a) Biskyrmions in the middle of a rectangle film. The sample is defined as $1~\upmu\mathrm{m}\times 0.5~\upmu\mathrm{m} \times160~\mathrm{nm}$ size with cells of $4~\mathrm{nm}\times 4~\mathrm{nm} \times4~\mathrm{nm}$. The magnetic easy axis lies in yz-plane. (b) The dependence of the topological charge $ Q $ with perpendicular position of the left biskyrmion in (a) during the process of removing $B_z$. (c) The variations of different energy terms $\Delta_E$ during the process of removing $B_z$. The initial values at $B_z$ = 0.14 T are subtracted from the energy terms for better illustration. $E_Z,~E_A,~E_X$, and $E_D$ denote Zeeman, anisotropy, exchange, and magnetostatic energy, respectively. (d)(e) The distribution of magnetostatic energy density $e_D$ on the contour surface of $ \mathbf{m}_z=0 $ at $B_z$ = 0.14 T and 0 T, respectively. Grey dots depict the location of BPs. }\label{fig:stripe}
\end{figure}

We have demonstrated that the relative position of BPs is the key factor to form biskyrmions. The remaining problem is why the BPs' locations are different between biskyrmions and bubbles. The residing area of the biskyrmion \cite{SM} inspires us to investigate the effect of the film edges. Therefore, we try to create biskyrmions in geometry-confined films, which happen to be the common case in application. As a result, two biskyrmions are generated in a rectangle film under zero field with the easy axis tilted at an angle of $ 15^\circ $, displayed in Fig.\ref{fig:stripe}(a). 
 
It turns out that type-I bubbles are favorable in confined geometry. Therefore, a y-component field is introduced to produce type-II bubbles at first. In detail, these bubbles are realized by relaxing the randomly magnetized state at an external magnetic field of (0, 0.05, 0.14) $\mathrm{T}$. Then the y-component and z-component are removed with steps of 0.01 $\mathrm{T}$ sequentially, and the state is relaxed in every step. After removing the y-component field, the elimination of the z-component field $B_z$ makes bubbles enlarge and transform to biskyrmions under the influence of edges. Fig.\ref{fig:stripe}(b) shows the variation of $Q$ versus $z$ during removing $B_z$, demonstrating the process of biskyrmion transition. Also, relaxing the bubble state under zero field directly can obtain quite similar structures \cite{SM}, indicating that the history of field decreasing has little influence on the formation of biskyrmions. In contrast, eliminating the magnetic field below a certain threshold value makes bubbles and biskyrmions turn into stripe domains if the tilted angle is large as in Fig.\ref{fig:sim}. 

To clarify the mechanism, we inspect the transition process of the left biskyrmion in Fig.\ref{fig:stripe}(a) during removing $B_z$. As can be seen in Fig.\ref{fig:stripe}(b), the variation of $Q$ versus $z$ indicates the motion of BPs during this process. To address the derivation of BPs' motion, the variations of energy terms are presented in Fig.\ref{fig:stripe}(c). The anisotropy energy $E_A$ and exchange energy $E_X$ only increase slightly. The change of $B_z$ makes the Zeeman energy $E_Z$ increase a lot, while the magnetostatic energy $E_D$ is the only decreasing term. Apart from $E_Z$, the total energy is reduced via $E_D$. Therefore, it can be concluded that the magnetostatic energy dominates the transition process. Furthermore, the distribution of magnetostatic energy density $e_D$ on the biskyrmion domain wall shown in Fig.\ref{fig:stripe}(d) and Fig.\ref{fig:stripe}(e) reveals the relation between BPs and $E_D$. Specifically, the energy distribution above the BP is larger than the one below the BP in Fig.\ref{fig:stripe}(d). As $B_z$ decreases, the BP is driven up to reduce the magnetostatic energy as in Fig.\ref{fig:stripe}(e), accompanied by the enlargement of the biskyrmion. This mechanism can also account for the observed edge effect, which can be attributed to the influence of the demagnetizing field. On the other hand, magnetic impurities may also trigger the formation of biskyrmions via influencing BPs.

In terms of experiments, THE may still be a convenient approach to confirm our findings since reconstructing 3D magnetic structures is technologically difficult. In detail, the dependence of THE with the external field can be observed while the number of bubbles remains constant. Quantitative experiments can be performed with stripe magnets to confirm the existence of biskyrmions.

In conclusion, we demonstrate the 3D structure of biskyrmions and analyze its formation mechanism based on micromagnetic simulation. Biskyrmions and type-II bubbles can form in a magnetic film with tilted uniaxial anisotropy and produce similar images under Lorentz TEM. In fact, biskyrmions and bubbles share a unified 3D structure, in which the relative position of two intrinsic BPs dominates the topological charge in the film middle. We also generate biskyrmions in a stripe magnet under zero field. By inspecting the process of biskyrmion formation, BPs are found to be driven by film edges via the demagnetizing field. Therefore, it can be assumed that biskyrmions can be generated in many materials because our model only requires geometry-confined ferromagnetic films with tilted uniaxial anisotropy.

The 3D model in this work may inspire both fundamental research and application development. The capability of encoding two states in the topological charges of a unified 3D object can lead to new architectures of data storage devices and other spintronics applications. On the other hand, this work demonstrates the 3D magnetization structure of magnetic bubbles under the influence of tilted magnetic anisotropy. This 3D object is the same class of particle-like excitations as the chiral bobber \cite{rybakov_new_2015,zheng_experimental_2018}, composed of a magnetization field and magnetic singularities. These results draw attention to the 3D aspect of magnetic nanostructures, opening a new avenue to study magnetic skyrmions.
\\

The data that support the findings of this study are available from the corresponding author upon reasonable request.
\\

    This work was supported by the National Key R\&D Program of China (Grant No. 2018YFA0306600, No. 2018YFF01012500), the National Natural Science Foundation of China (Grant No. 81788101, No. 11874338, No. T2125011), the Chinese Academy of Sciences (Grants No. XDC07000000, No. GJJSTD20200001, No. QYZDY-SSW-SLH004, No. KJZD-SW-M01, No. ZDZBGCH2021002),  Anhui Initiative in Quantum Information Technologies (Grant No. AHY050000), Natural Science Foundation of Anhui (Grant No. 1808085J09) and the Fundamental Research Funds for the Central Universities. Y.Z. acknowledges the support by Guangdong Special Support Project (2019BT02X030), Shenzhen Fundamental Research Fund (Grant No. JCYJ20210324120213037), Shenzhen Peacock Group Plan (KQTD20180413181702403), Pearl River Recruitment Program of Talents (2017GC010293) and National Natural Science Foundation of China (11974298, 61961136006).

\bibliography{paper}

\end{document}